\documentclass[10pt, conference]{IEEEtran}
\usepackage{silence}
\IEEEoverridecommandlockouts

\usepackage{cite}
\usepackage{amsmath,amssymb,amsfonts}
\usepackage{float} 
\usepackage{hyperref}
\usepackage{algorithm,algpseudocode}
\usepackage{graphicx}
\usepackage{textcomp}
\usepackage[dvipsnames]{xcolor}
\usepackage{verbatim}
\usepackage{lipsum}  
\usepackage{balance}
\usepackage{caption}

\def\BibTeX{{\rm B\kern-.05em{\sc i\kern-.025em b}\kern-.08em
    T\kern-.1667em\lower.7ex\hbox{E}\kern-.125emX}}
    
\begin{document}

\bibliographystyle{IEEEtran}

\title{Measuring Improvement of F$_1$-Scores in Detection of Self-Admitted Technical Debt}

\author{
    \IEEEauthorblockN{William Aiken\normalfont\textsuperscript{§}\thanks{§ Joint first authors},
                      Paul K. Mvula\normalfont\textsuperscript{§},
                      Paula Branco,
                      Guy-Vincent Jourdan,
                      Mehrdad Sabetzadeh, and 
                      Herna Viktor}

\IEEEauthorblockA{School of Electrical Engineering and Computer Science (EECS)\\
University of Ottawa,
Ottawa, ON, Canada.\\
Emails: \{waike081,
pmvul089,
pbranco,
gjourdan,
m.sabetzadeh, hviktor\}@uottawa.ca}}

\maketitle

\begin{abstract}

Artificial Intelligence and Machine Learning have witnessed rapid, significant improvements in Natural Language Processing (NLP) tasks. Utilizing Deep Learning, researchers have taken advantage of repository comments in Software Engineering to produce accurate methods for detecting Self-Admitted Technical Debt (SATD) from 20 open-source Java projects' code. In this work, we improve SATD detection with a novel approach that leverages the Bidirectional Encoder Representations from Transformers (BERT) architecture. For comparison, we re-evaluated previous deep learning methods and applied stratified 10-fold cross-validation to report reliable F$_1$-scores. We examine our model in both cross-project and intra-project contexts. For each context, we use re-sampling and duplication as augmentation strategies to account for data imbalance. We find that our trained BERT model improves over the best performance of all previous methods in 19 of the 20 projects in cross-project scenarios. However, the data augmentation techniques were not sufficient to overcome the lack of data present in the intra-project scenarios, and existing methods still perform better. Future research will look into ways to diversify SATD datasets in order to maximize the latent power in large BERT models.

\end{abstract}


\section{Introduction}

Technical debt is the phenomenon where engineers adopt a limited or easy solution at  implementation instead of a more efficient but time-consuming one, usually in order to accelerate the delivery of features or software releases~\cite{cunningham1992wycash}. Technical debt that has been knowingly introduced in the form of source code comments during software development is referred to as Self-Admitted Technical Debt (SATD)~\cite{potdar2014exploratory}. A recent survey~\cite{vishal_tech_debt} reported that technical debt accounts for 20\% to 40\% of the value of software projects before depreciation and increases maintenance costs over time. Therefore, detecting SATD in source code in a timely manner can allow proper management and mitigation by software engineers and decision-makers. 

Various approaches have been proposed to detect SATD in source code comments~\cite{potdar2014exploratory, maldonado2017using, huang2018identifying, ren_neural_2019, yu_identifying_2022}. However, in all cases, the authors confront a recurring issue: SATD comments represent an extremely imbalanced task of classifying the minority SATD vs. the majority non-SATD comments. Among the datasets leveraged~\cite{guo_how_2021}, SATD comments make constitute the highest percentage in \texttt{ArgoUML} at 17.86\%, and the lowest amount in \texttt{SpringFramework} at 1.27\% of the comments. In intra-project train-test scenarios, significant issues can arise. Assuming a dataset has 100 SATD comments, a 90-10 train-test split would result in only 10 SATD comments for evaluation, a situation which occurs in work in this field. 

Further exacerbating the situation, while Ren et al.~\cite{ren_neural_2019} have explored the significant correlation of textual patterns with SATD comments (e.g. ``todo'', ``hack'', etc.), little of the existing literature accounts for the large differences in difficulty that each random train-test split may have. An example of this disparity is shown in Table~\ref{table: easy-difficult}. In order to account for potential selection bias in intra-project scenarios, one would expect to see a stratified K-fold cross-validation framework for empirical assessment. However, this technique is missing in some recent research in this field~\cite{ren_neural_2019}. As a result, the current benchmark of classification performance of state-of-the-art neural network architectures on this task may be unclear. The contributions of this work are as follows:

\begingroup
\begin{itemize}
    \item We implement a neural network based on architectures designed for Natural Language Processing (NLP) tasks; specifically, we leverage the Bidirectional Encoder Representations from Transformers (BERT)~\cite{devlin_bert_2019} architecture. We use this model in both cross- and intra-project settings and evaluate its F$_1$ performance using a robust stratified 10-fold cross-validation framework.
    \item We re-evaluate the work of Ren et al.~\cite{ren_neural_2019} which uses a Convolutional Neural Network (CNN) approach for the same task using the same stratified 10-fold validation for comparison, which was previously lacking. 
    \item For the data-impoverished intra-project scenarios, we further apply custom minority re-sampling and duplication data augmentation techniques in an attempt to improve the performance of the BERT model against the imbalance of the SATD vs. non-SATD comments.
\end{itemize} 
\endgroup

We find that our BERT model performs with the highest F$_1$-score, but only in cross-project scenarios, where there is sufficient training data. The intra-project scenarios yield overfit results, even when using our data augmentation techniques. Future work must continue to focus on better data augmentation for text-based domains in order to fully leverage large models such as BERT in data-limited scenarios. Further information about the datasets and the source code required to reproduce our results are available at ~\href{https://doi.org/10.5281/zenodo.7697129}{10.5281/zenodo.7697129}.

\begin{table}[t!]
\captionsetup{font=footnotesize}
\scriptsize
\caption{SATD comments of varying difficulty from the ApacheAnt}
\label{table: easy-difficult}
\begin{tabular}{l}
\hline
\textbf{``Easy'' SATD Comments (contain clear trigger words)}                                                                                                              \\ \hline
//TODO: nothing appears to read this but is set using a public setter.                                                                       \\
// MAC OS 9 and previous //TODO: I have no idea how to get it...\\

\hline
\textbf{``Hard'' SATD Comments (do not contain clear trigger words)}                                                                                                              \\ \hline
// sorry - otherwise we will get a ClassCastException because the MockCache...\\
//these are pathological cases, but retained in case somebody //subclassed us.                                                               \\
\hline
\end{tabular}
\end{table}

\section{Background}

Since Potdar et al.~\cite{potdar2014exploratory} conducted a study on detecting SATD from source code comments, several other methods have been proposed to detect SATD. Potdar et al. manually extracted 62 specific patterns to detect SATD using data from Eclipse, Chromium OS, ArgoUML and Apache HTTPd. Huang et al.~\cite{huang2018identifying} applied tokenization, stop-word removal, and stemming to the code comments; they then selected the most useful features using Information Gain. Although these two approaches performed well on the projects they were tested on, their main disadvantage is that they have limited generalizability and do not adapt well to cross-project settings because they rely on hand-crafted features that: (i) do not take into account word relationships, and (ii) do not effectively capture changes over time. Huang et al.~\cite{huang2018identifying} did, however, use 10-fold cross-validation framework, which provides confidence in their results. Instead of extracting fixed features, Maldonado et al.~\cite{maldonado2017using} used the maximum entropy classifier to extract features from input data and detect SATD automatically. 

Guo et al.~\cite{guo_how_2021} further explored both an approach that strictly or fuzzily ``Matches task Annotation Tags'' (MAT) to identify SATD comments. In another work, Ren et al.~\cite{ren_neural_2019} relied on neural networks for this classification. They first generated word embeddings using the continuous skip-gram model of Word2Vec and fed the embeddings to a CNN. The CNN incorporated a weighted cross-entropy loss whose objective function penalized more the wrong predictions on instances that belonged to the minority class than those that belonged to the majority class. Therefore, their cross-entropy loss attempted to overcome the data imbalance issue. 

More recent work has also explored classifying code comments for various ends in multiple languages, each with their own unique approach. In their study, Gao et al.~\cite{gao_automating_2021} aimed to identify ``\textit{obsolete \textbf{TODO}}'' comments within Java and Python software projects separately. These types of comments constitute only a subset of SATDs and can have negative effects on program comprehension, cause communication issues among developers, and create confusion for those working on future developments. However, the authors did not develop a 10-fold cross-validation framework to bolster the reliability and validity of their findings, nor did they evaluate their approach in cross-project settings. Pascarella et al.~\cite{pascarella2019classifying} took a more general approach to classifying code comments in Java, rather than focusing solely on SATD. They created an additional category called ``Under Development'', which had subcategories such as ``TODO'', ``Incomplete'', and ``Commented Code''. Although we did not use this category in our work, it provides another avenue for future research in comment classification tasks.


Sharma et al.~\cite{sharma_self_admitted_2022} proposed a method to detect SATDs in the R programming language. They applied several classifiers, including Max Entropy, SVMs, CNN, and two BERT-based models, and they reported that the latter outperformed the other models. The authors employed a cross-validation framework to evaluate their method, but they did not conduct intra- and cross-project analyses due to their designed lack of correlation between the comment and its project of origin. 

Finally, Prenner and Robbes~\cite{prenner_making_2021} evaluated the performance of Transformer models on small- and medium-sized software engineering datasets. Their proposed models, including StackOBERTflow, yielded the highest results in cross-project scenarios, outperforming both Ren et al.\cite{ren_neural_2019} and Maldonado et al.\cite{maldonado2017using}'s approaches on the 10 Java project comments annotated by Maldonado et al. However, the authors did not report intra-project results.

\subsection{Evaluation}

\begin{table}[t]
\captionsetup{font=footnotesize}
\centering
\scriptsize
\caption{Cross-Project F$_1$-Scores from Existing Work}
\label{table: cross-project results}
\begin{tabular}{|c|c|c|c|c|c|c|c|}
\hline
\multicolumn{1}{|c|}{\textbf{Dataset}} & \multicolumn{1}{c|}{\textbf{\shortstack{Huang \\ (TM)}}} & \multicolumn{1}{c|}{\textbf{\shortstack{Ren \\ (CNN)}}} & \multicolumn{1}{c|}{\textbf{\shortstack{Guo \\ (ext)}}} & \multicolumn{1}{c|}{\textbf{\shortstack{Yu \\ (Easy)}}} & \multicolumn{1}{c|}{\textbf{\shortstack{Yu \\ (JB)}}} & \multicolumn{1}{c|}{\textbf{\shortstack{Prenner \\ (BERT)}}} & \multicolumn{1}{c|}{\textbf{Best}} \\ \hline
\textit{ApacheAnt}                     & 0.51                                       & 0.66                           & 0.60                                        & 0.38                                          & 0.21                       &         \textbf{0.70}        &    0.70                              \\ \hline
\textit{ArgoUML}                       & 0.83                                       & 0.88                           & 0.87                                          & 0.87                                          & 0.76                     &          \textbf{0.90}          &  0.90                                \\ \hline
\textit{Columba}                       & 0.81                                       & 0.85                                    & 0.89                                          & 0.89                                 & 0.49                     &          \textbf{0.91}         &  0.91                                \\ \hline
\textit{EMF}                           & 0.54                                       & 0.68                                    & 0.72                                 & 0.44                                          & 0.19                      &         \textbf{0.73}         &      0.73                            \\ \hline
\textit{Hibernate}                     & 0.80                                       & 0.83                                    & 0.83                                 & 0.83                                          & 0.64                    &            \textbf{0.88}        &      0.88                            \\ \hline
\textit{JEdit}                         & 0.49                                       & 0.60                           & 0.54                                          & 0.35                                          & 0.27                     &            \textbf{0.73}       &      0.73                            \\ \hline
\textit{JFreeChart}                    & 0.68                                       & 0.74                           & 0.72                                          & 0.57                                          & 0.37                     &            \textbf{0.78}       &   0.78                               \\ \hline
\textit{JMeter}                        & \textbf{0.88}                              & 0.83                                    & 0.85                                          & 0.80                                          & 0.23                      &           0.87       &    0.88                              \\ \hline
\textit{JRuby}                         & 0.80                                       & 0.86                                    & 0.89                                 & 0.67                                          & 0.77                     &            \textbf{0.91}       &    0.91                              \\ \hline
\textit{SQuirrel}                      & 0.68                                       & 0.74                                    & 0.76                                 & 0.66                                          & 0.35                      &           \textbf{0.79}       &   0.79                               \\ \hline\hline
Average                                & 0.70                                       & 0.76                                    & 0.77                                 & 0.64                                          & 0.43                       &           \textbf{0.82}      &   0.82                               \\ \hline
\end{tabular}
\vspace{-0.5em}
\end{table}

\begin{table}[t!]
\captionsetup{font=footnotesize}
\scriptsize
\centering
\caption{Intra-Project F$_1$-Scores from Existing Work}
\label{table: intra-project results}
\begin{tabular}{|l|c|c|c|c|}
\hline
\multicolumn{1}{|c|}{\textbf{Dataset}} & \multicolumn{1}{c|}{\textbf{\shortstack{Huang \\ (TM)}}} & \multicolumn{1}{c|}{\textbf{\shortstack{Ren Reported \\ (CNN)}}} & \multicolumn{1}{c|}{\textbf{\shortstack{Ren 10-fold \\ (CNN)}}} & \multicolumn{1}{c|}{\textbf{Best}} \\ 
    \hline
    \textit{ApacheAnt} & 0.653 & 0.445 & \textbf{0.66} & 0.66 \\ \hline
    \textit{ArgoUML} & 0.618 & \textbf{0.932} & 0.92 & 0.932 \\ \hline
    \textit{Columba} & 0.726 & 0.741 & \textbf{0.83} & 0.83 \\ \hline
    \textit{EMF} & - & 0.532 & \textbf{0.64} & 0.64 \\ \hline
    \textit{Hibernate} & 0.726 & 0.887 & \textbf{0.89} & 0.89 \\ \hline
    \textit{JEdit} & 0.617 & 0.622 & \textbf{0.74} & 0.74 \\ \hline
    \textit{JFreeChart} & 0.47 & \textbf{0.795} & 0.68 & 0.795 \\ \hline
    \textit{JMeter} & 0.728 & 0.867 & \textbf{0.9} & 0.9 \\ \hline
    \textit{JRuby} & 0.749 & 0.881 & \textbf{0.91} & 0.91 \\ \hline
    \textit{SQuirrel} & 0.548 & 0.813 & \textbf{0.84} & 0.84 \\ \hline
    \hline
    \textbf{Average} & 0.648 & 0.752 & \textbf{0.801} & 0.801 \\ \hline
\end{tabular}
\vspace{-\baselineskip}
\end{table}

As a baseline in the \textit{intra-project} setting, we primarily refer to Ren et al.~\cite{ren_neural_2019} because it is the most similar in task and approach, and to the best of our knowledge, is the state-of-the-art in the domain of classifying SATD \textit{without} human intervention within the same project. Specifically, Ren et al.~\cite{ren_neural_2019} used a CNN with word-embedding dimension set at 300, 6 different filter window sizes (1- to 6-grams), and 128 filters for each window size.  Their convolutional layer is followed by a pooling layer, a fully connected layer (with dropout), and a Softmax classification layer. However, there are some cases where the approaches by Huang et al.~\cite{huang2018identifying}, Guo et al.~\cite{guo_how_2021}, Yu et al.~\cite{yu_identifying_2022} and Prenner and Robbes~\cite{prenner_making_2021} are superior. We include their results in the tables as well when applicable.

In the intra-project scenarios, we re-calculated the F$_1$-scores via stratified 10-fold cross-validation where necessary for previous work. In a default 90-10 train-test split, we noticed repeated fluctuations in performance both in our work and in previous work. We posit this occurs due to the presence or lack of specific key words (``ugly'', ``TODO'', etc.) in the very small intra-project test set, which we represent in Table~\ref{table: easy-difficult}.

Note, however, that the columns for the text mining (TM) approach by Huang in Table~\ref{table: cross-project results} are those presented in Ren et al.'s work~\cite{ren_neural_2019} to reflect the \texttt{Dataset-M} training data. Additionally, while the work by Guo et al.~\cite{guo_how_2021} attempted to recreate the cross-project results by Ren et al.~\cite{ren_neural_2019} for similar evaluation of the progress on this task, they reported that they were unable to reproduce the results. Although we did not attempt to verify the \textit{cross-project} results, in our case, we were able to reproduce the \textit{intra-project} results of Ren et al. from the code in Guo et al.'s repository (which they received from Ren et al.), but only after applying K-fold cross-validation. In fact, as shown in Table~\ref{table: intra-project results}, we obtained \textit{higher} intra-project results in 8 out of the 10 datasets than reported in the original paper~\cite{ren_neural_2019}.  Huang et al.~\cite{huang2018identifying}'s cross-validated intra-project results are also included in Table~\ref{table: intra-project results}.

\section{Our Approach}\label{AA}

\subsection{Datasets}

As previously established in the work by Guo et al.~\cite{guo_how_2021}, we refer to the initial batch of 10 annotated projects' comments as \texttt{Dataset-M}, which includes ApacheAnt, ArgoUML, Columba, EMF, Hibernate, JEdit, JFreeChart, JMeter, JRuby, and SQuirrel. Moreover, we also follow the convention of referring to the second batch of 10 annotated projects' comments as \texttt{Dataset-G}, which includes Dubbo, Gradle, Groovy, Hive, Maven, Poi, SpringFramework, Storm, Tomcat, and Zookeeper.

For the cross-project scenarios, we follow the trend also established in previous work in using hold-one-out validation as a method to train a classifier to detect the SATD comments. In other words, we use 19 repositories' comments as training data, and the remaining 1 repository's comments as test data. In Guo et al.~\cite{guo_how_2021}, this is referred to as an ``MTO'' scenario (short for many-to-one). Although we refer to this scenario as cross-project in our work, the train-split set up is identical to that of Guo et al.~\cite{guo_how_2021}.

Because comments are primarily in standard English augmented by code-like information, we use the \texttt{bert-base-cased} model originally proposed in Devlin et al.~\cite{devlin_bert_2019}. Comments often contain casual writing styles (e.g. \textit{``// FRICKIN' HACK!!!!! For some reason, deleting a string at offset 0   does not get done properly, so first replace and remove after parsing}''), so we decided on a case-sensitive model due to the frequent use of capitalization for emphasis.

\subsection{Tokenization and Pre-processing}
\captionsetup{font=small}
\label{subsec: tokenization}


Many comments include code-like contents particularly when the author refers to specific parts of the code within the comment, such as: ``// ClassLoader parentLoader = Thread.currentThread().getContextClassLoader();''. Because our BERT model and tokenizer is not designed to handle raw code, we preprocess the code according to Java's best practices for variable names. In other words, ``new CharParserForJavaOrSomething();'' would be pre-processed to become ``new Char Parser For Java Or Something();''. This was the only pre-processing technique applied before entering the rest of the pipeline.

We also leveraged the same base BERT tokenizer corresponding to \texttt{bert-base-cased}, which already contains 28,996 tokens. In order to fine-tune a BERT model on a specific domain, it is required to augment the vocabulary of the model tokenizer with additional words that should represent tokens for the domain in question. In our approach, we wanted to have one universal tokenizer usable across all configurations. To achieve this, we identified all words that appeared within the dataset that did not already exist as tokens within the default \texttt{bert-base-cased} tokenizer. From these, we then made a comprehensive list of all potential tokens that occur in greater than 25\% of the source-code repositories. The list also included the following tokens because they also reached that same appearance criteria : ``/*, */, //, {[]}, (), and ;''. Additionally, we aimed the model to be able to differentiate comment-related indicators ``// this is a comment'' from standard English usages such as ``and/or'', function call endings ``function()'' from standard English parenthetical phrases, etc. Out of the context of indicating comments, these symbols can have other meanings (e.g., ``*'' indicate emphasis when surrounding a word(s)) that we also wanted to avoid.

The full list results in 1,771 new tokens; however, due to the mixture of natural language and code, many of these are not valid or productive word presentations. Such flawed extractions include (e.g., ``ns'', ``2.2''). While some short words such as ``li'' could have special significance, their inclusion would cause unwarranted tokenization in other contexts. We removed such tokens. After manual verification of all tokens, requiring approximately 3 hours, we were left with 1,653 final, meaningful tokens.

\subsection{Scenarios Tested}

We included two scenarios for this experiment: a cross-project scenario, and an intra-project scenario. The cross-project setup is designed so that the BERT classification model is trained on the annotated comments from 19 projects, and the remaining 1 project is the test set. The 20 projects result from the combination of \texttt{Dataset-M} and \texttt{Dataset-G}. In the intra-project scenarios, we use 90\% of the data from \textit{within one project} to train the BERT classifier, and the remaining 10\% of that repository is the test set. Because of the discrepancies presented in Table~\ref{table: easy-difficult}, we perform the intra-project scenarios in a stratified 10-fold cross-validation framework.

We argue that both scenarios yield valuable insights. An analysis of the cross-project scenario demonstrates the performance a software engineer could reasonably expect to receive on a \textit{novel repository} that they are analyzing. On the other hand, the intra-project scenario could be said to represent the performance a software engineer could reasonably expect on a \textit{similar yet novel annotation task} within their repository. While the current binary approach of ``SATD'' vs. ``non-SATD'' is valuable, a team may take an alternative classification task. They may only be interested in annotating their own software repository for an alternate, yet similar, classification task. As a result, we include the intra-project performance on the ``SATD'' vs. ``non-SATD'' classification task as a representation of such a limited data scenario.

\subsection{Approaches}

\textbf{Baseline Approach.} While in Ren et al.'s CNN approach~\cite{ren_neural_2019}, the authors weighted the loss of the neural network to encourage the model to give more weight to potential SATD comments, we opted to avoid weighting positive examples higher as we also wanted to provide data augmentation schemes to account for this difference. Our baseline approach contains no change in the loss of the model's predictions in the data loading pipeline, no change to the data loader, and no change to the training data.

\textbf{Forced Minority Re-Sampling.} To account for the highly imbalanced nature of the datasets, our initial approach was to simply get the positive examples into the training pipeline more often. To allow for a more dynamic nature, we re-sampled the positive examples at custom rates. Specifically, during training, for 10\% of training batches during an epoch, we would also sample from the minority (SATD) class if the ratio was below 3:1 (non-SATD to SATD) for that batch. These values were set by empirical observations and may not be optimal for this domain. We refer to this technique as Forced Minority Re-sampling (FMR) shown in Algorithm~\ref{algo: forced-minority-resampling}. While oversampling minority classes is a common approach in imbalanced tasks, it may result in limitations such as oversampling uninformative or noisy examples, issues our method does not currently address.

\begin{figure}[t]
    \vspace{-1.5em}
    \begin{algorithm}[H]
    \scriptsize
    \captionsetup{font=scriptsize}
    \caption{Forced Minority Re-sampling (FMR)}
    \label{algo: forced-minority-resampling}
    \begin{algorithmic}[1]
     \Require Minority class samples (SATD comments)
     \Ensure  Batch that contains both SATD and non-SATD comments
    \State Compile SATD comments into one text corpus.
    \State During training loop, $n$\% of the time, draw from SATD set until the ratio of non-SATD to SATD reaches desired level $\rho$ (our work: $n=10\%; \rho=3:1$).
    \State Repeat until model convergence criteria.
    \end{algorithmic}
    \end{algorithm}
    \vspace{-1.5em}
\end{figure}

\begin{figure}
    \vspace{-1.5em}
    \begin{algorithm}[H]
    \scriptsize
    \captionsetup{font=scriptsize}
    \caption{Duplicated SATD FMR (DUP)}
    \label{algo: dup-fmr}
    \begin{algorithmic}[1]
     \Require Minority class samples (SATD comments)
     \Require Pre-defined key trigger words
     \Ensure  Batch that contains both SATD and non-SATD comments
    \State Compile SATD comments into one text corpus.
    \State Duplicate all SATD comments in train set
    \State For all duplicates, remove key trigger words
    \State During training loop, $n$\% of the time, draw from SATD set until the ratio of non-SATD to SATD reaches desired level $\rho$ (our work: $n=10\%; \rho=3:1$).
    \State Repeat until model convergence criteria.
    \end{algorithmic}
    \end{algorithm}
    \vspace{-2.5em}
\end{figure}

\textbf{Data Augmentation.} Because of the plethora of comments with keywords that clearly indicate SATD, we explored the ability to create duplicates of these comments simply with those keywords removed. This duplication approach was done on top of the FMR approach to determine if adding data augmentation to the comments would positively impact the performance of the FMR approach. We refer to this technique as Duplicated SATD FMR (DUP) shown in Algorithm~\ref{algo: dup-fmr}. Note that this duplication was done on the ``easy'' comments, i.e., only those that contained trigger words. For example, ``// FIXME: This should probably...'' would generate an additional entry ``// This should probably...''.

\begin{table}[t!]
\captionsetup{font=footnotesize}
\scriptsize
    \centering
    \caption{Our 19-to-1 Cross-Project F$_1$-Scores of BERT Model with Existing Best (for Projects in \texttt{Dataset-M})}
    \label{table: our cross-results dataset-m}
    \begin{tabular}{|l|l|l|l|c|}
    \hline
    \multicolumn{1}{|c|}{\textbf{Project}} & \multicolumn{1}{c|}{\textbf{Baseline}} & \multicolumn{1}{c|}{\textbf{FMR}} & \multicolumn{1}{c|}{\textbf{DUP}} & \textbf{\shortstack{Existing \\ Best}} \\ \hline
    \textit{ApacheAnt}                     & 0.492                                  & 0.804                             & \textbf{0.807}                    & 0.660                  \\ \hline
    \textit{ArgoUML}                       & 0.917                                  & 0.921                             & \textbf{0.925}                    & 0.878                  \\ \hline
    \textit{Columba}                       & 0.893                                  & 0.850                             & 0.848                             & \textbf{0.890}         \\ \hline
    \textit{EMF}                           & 0.493                                  & 0.745                             & \textbf{0.782}                    & 0.715                  \\ \hline
    \textit{Hibernate}                     & 0.897                                  & \textbf{0.898}                    & 0.868                             & 0.831                  \\ \hline
    \textit{JEdit}                         & 0.709                                  & \textbf{0.767}                    & 0.757                             & 0.599                  \\ \hline
    \textit{JFreeChart}                    & 0.490                                  & \textbf{0.879}                    & 0.872                             & 0.739                  \\ \hline
    \textit{JMeter}                        & 0.912                                  & \textbf{0.923}                    & 0.913                             & 0.881                  \\ \hline
    \textit{JRuby}                         & \textbf{0.938}                         & 0.922                             & 0.472                             & 0.897                  \\ \hline
    \textit{SQuirrel}                      & 0.865                                  & \textbf{0.869}                    & 0.852                             & 0.766                  \\ \hline\hline
    \textit{Average}                       & 0.760                                  & \textbf{0.858}                    & 0.810                             & 0.768                  \\ \hline
    \end{tabular}

\end{table}

\begin{table}[t!]
\captionsetup{font=footnotesize}
\scriptsize
    \centering
    \caption{Our 19-to-1 Cross-Project F$_1$-Scores of BERT Model with Existing Best (for Projects in \texttt{Dataset-G})}
    \label{table: our cross-results dataset-g}
    \begin{tabular}{|l|l|l|l|c|}
    \hline
    \multicolumn{1}{|c|}{\textbf{Project}} & \multicolumn{1}{c|}{\textbf{Baseline}} & \multicolumn{1}{c|}{\textbf{FMR}} & \multicolumn{1}{c|}{\textbf{DUP}} & \multicolumn{1}{c|}{\textbf{\shortstack{Guo et al. \\ Best}}} \\ \hline
    \textit{Dubbo}                         & 0.856                                  & \textbf{0.880}                    & 0.487                             & 0.737                                     \\ \hline
    \textit{Gradle}                        & 0.853                                  & 0.852                             & \textbf{0.866}                    & 0.703                                     \\ \hline
    \textit{Groovy}                        & \textbf{0.892}                         & 0.890                             & 0.486                             & 0.782                                     \\ \hline
    \textit{Hive}                          & \textbf{0.884}                         & 0.869                             & 0.882                             & 0.789                                     \\ \hline
    \textit{Maven}                         & \textbf{0.867}                         & 0.859                             & 0.861                             & 0.718                                     \\ \hline
    \textit{Poi}                           & 0.871                                  & \textbf{0.905}                    & 0.868                             & 0.850                                     \\ \hline
    \textit{Spring}                        & 0.834                                  & \textbf{0.847}                    & 0.843                             & 0.673                                     \\ \hline
    \textit{Storm}                         & 0.862                                  & 0.872                             & \textbf{0.876}                    & 0.709                                     \\ \hline
    \textit{Tomcat}                        & 0.494                                  & 0.863                             & \textbf{0.872}                    & 0.763                                     \\ \hline
    \textit{Zookeeper}                     & \textbf{0.857}                         & 0.840                             & 0.848                             & 0.617                                     \\ \hline\hline
    \textit{Average}                       & 0.827                                  & \textbf{0.868}                    & 0.789                             & 0.734                                     \\ \hline
    \end{tabular}
\end{table}

\begin{table}[t]
\captionsetup{font=footnotesize}
\scriptsize
    \centering
    \caption{Intra-Project F$_1$-Scores of 10-fold CV of BERT Models (for Projects in \texttt{Dataset-M})}
    \label{table: our intra-results dataset-m}
    \begin{tabular}{|c|c|c|c|c|}
    \hline
    \textbf{Project} & \textbf{Baseline} & \textbf{FMR} & \textbf{DUP} & \textbf{\shortstack{Existing \\ Best}} \\ \hline
    \textit{ApacheAnt}        & 0.654             & 0.688        & \textbf{0.663}        & 0.66         \\ \hline
    \textit{ArgoUML}          & 0.729             & 0.777        & 0.745        & \textbf{0.932}         \\ \hline
    \textit{Columba}          & \textbf{0.876}             & 0.834        & 0.814        & 0.83         \\ \hline
    \textit{EMF}              & \textbf{0.801}             & 0.799        & 0.746        & 0.64         \\ \hline
    \textit{Hibernate}        & 0.804             & 0.78         & 0.765        & \textbf{0.89}         \\ \hline
    \textit{JEdit}            & 0.755             & 0.737        & \textbf{0.759}        & 0.74         \\ \hline
    \textit{JFreeChart}       & 0.681             & 0.695        & 0.707        & \textbf{0.795}         \\ \hline
    \textit{JMeter}           & 0.874             & 0.876        & 0.877        & \textbf{0.9}         \\ \hline
    \textit{JRuby}            & 0.864             & 0.839        & 0.841        & \textbf{0.91}         \\ \hline
    \textit{SQuirrel}         & 0.864             & 0.849        & 0.842        & 0.84         \\ \hline\hline
    \textit{Average}          & 0.790             & 0.787        & 0.776        & \textbf{0.801}         \\ \hline
    \end{tabular}
    \vspace{-0.5em}
\end{table}

\begin{table}[t]
\captionsetup{font=footnotesize}
\scriptsize
\centering
    \caption{Intra-Project F$_1$-Scores of 10-fold CV of BERT Models (for Projects in \texttt{Dataset-G})}
    \label{table: our intra-results dataset-g}
    \begin{tabular}{|l|c|c|c|}
    \hline
    \multicolumn{1}{|c|}{\textbf{Project}} & \multicolumn{1}{c|}{\textbf{Baseline}} & \multicolumn{1}{c|}{\textbf{FMR}} & \multicolumn{1}{c|}{\textbf{DUP}} \\ \hline
    \textit{Dubbo}                                  & 0.661                                  & 0.707                             & \textbf{0.716}                    \\ \hline
    \textit{Gradle}                                 & \textbf{0.777}                         & 0.761                             & 0.727                             \\ \hline
    \textit{Groovy}                                 & 0.812                                  & 0.789                             & \textbf{0.816}                    \\ \hline
    \textit{Hive}                                   & 0.785                                  & 0.770                             & \textbf{0.793}                    \\ \hline
    \textit{Maven}                                  & \textbf{0.767}                         & 0.748                             & 0.737                             \\ \hline
    \textit{Poi}                                    & 0.877                                  & \textbf{0.921}                    & 0.879                             \\ \hline
    \textit{Spring}                        & \textbf{0.757}                         & 0.743                             & 0.773                             \\ \hline
    \textit{Storm}                                  & 0.747                                  & \textbf{0.778}                    & 0.756                             \\ \hline
    \textit{Tomcat}                                 & 0.823                                  & 0.791                             & \textbf{0.823}                    \\ \hline
    \textit{Zookeeper}                              & \textbf{0.705}                         & 0.698                             & 0.695                             \\ \hline\hline
    \textit{Average}                                & 0.771                                  & 0.770                             & \textbf{0.771}                    \\ \hline
    \end{tabular}
\vspace{-1.5em}
\end{table}

\section{Results}

Our results offer a look at the improvement of leveraging a BERT model for classification of SATD comments in source code repositories. For the cross-project scenarios, we record an improvement in all but one project in the \texttt{Dataset-M} collection compared to the best of all existing approaches per project, shown in Table~\ref{table: our cross-results dataset-m}. Similarly, for the \texttt{Dataset-G} collection, we find that our model outperforms existing work; however, we can only compare with the reported results by Guo et al.~\cite{guo_how_2021} because earlier work did not include this collection. The results under Guo et al. are their best results out of their various pre-existing reported scenarios (MAT or MAT-ext with either strict or fuzzy tag matching) in Table~\ref{table: our cross-results dataset-g}.

None of our BERT-based approaches (baseline, FMR, or DUP augmentation) were able to match the existing performances on the intra-project scenario on average, as shown in Table~\ref{table: our intra-results dataset-m}. Other work continues to report that the size of the dataset plays an important role and that datasets of this size are still considered too small to train a transfer-learning algorithm that generalizes well to unseen data~\cite{prenner_making_2021}. To our knowledge, no previous work has performed intra-project evaluation on the projects in \texttt{Dataset-G}; our stratified 10-fold cross-validated results for this projects are presented in Table~\ref{table: our intra-results dataset-g}. We conclude that future work must continue to focus on better data augmentation techniques for text-based domains in order to fully leverage of large models such as BERT in data-poor scenarios.

\section{Future Work}

Going forward, the SATD classification task would benefit from further explicit exploration of the ``easy'' vs. ``hard'' distinction of these comments, as shown in Table~\ref{table: easy-difficult}. Specifically, we plan to leverage Jitterbug's open-sourced detection proposal~\cite{JitterBugGitHub} across all datasets in the repository~\cite{NapluesGithub} to give clearer ``easy''/``hard'' delineations for future models' evaluation. We aim to utilize this additional information to improve our FMR approach by re-sampling more insightful, difficult-to-classify comments. Moreover, similar annotation work by Pascarella et al.~\cite{pascarella2019classifying} also includes ``Under Development'' labels of ``TODO'' and ``Incomplete'' comments, which could serve as an additional SATD dataset on which to evaluate.

We advocate for additional data augmentation techniques in NLP, as the duplication method we tested only yielded minor improvement in some instances. We plan to explore alternative methods, such as using BERT models for synonym replacement in under-represented cases, incorporating LMOTE~\cite{leekha_multi-task_2020}, and TLMOTE~\cite{choudhry_tlmote_2022}, a variant of LMOTE where three language models are trained on 4, 5, and 6-grams.

Finally, there are some limitations to leveraging BERT models. Specifically, some manual effort is required in the tokenization process, and large BERT models require significant training time. Future work should explore automated methods for reducing tokenization effort and deciding model size.

\section{Conclusion}

In conclusion, NLP pipelines that leverage large BERT models such as \texttt{bert-base-cased} are able to improve state-of-the-art performance of SATD-classification tasks, but only in scenarios where ample data is present, such as when the comments from multiple repositories are available. In intra-project scenarios, where only the data from one project is available, the models appear to heavily overfit, and they fall short of existing approaches, even those of non-NLP-specific neural network architectures. We explored data re-sampling techniques and a duplication technique of SATD comments where SATD-specific keywords were removed in an attempt to discourage overfitting. These efforts, however, were insufficient to overcome this constraint. 

\section{Acknowledgements}

We thank the anonymous reviewers for their constructive comments. This work was supported by the Natural Sciences and Engineering Research Council of Canada, the Vector Institute, and The IBM Center for Advanced Studies (CAS) Canada within Project 1059. We are also grateful to the Digital Research Alliance of Canada (the Alliance) for access to their High-Performance Computing clusters.

\bibliography{references.bib}

\end{document}